\documentstyle[epsfig,draft]{elsart}              

\begin{document}
\begin{frontmatter}
                                                                               
\title{An improved Metropolis algorithm for hard core systems} 

\author{A.~Jaster}

\address{Universit\"{a}t - GH Siegen, D-57068 Siegen, Germany}
                                                                                
             
\maketitle
\begin{abstract}

We present an improved Metropolis algorithm  for arbitrary hard 
core systems in any dimensions. In the new updating scheme 
the conventional Metropolis step of a single particle
is replaced by a
collective step of a chain of particles.
For the two-dimensional hard sphere model 
we show that  this algorithm essentially reduces
autocorrelation times in the vicinity of the transition point.
\end{abstract}


\end{frontmatter}

\section{Introduction}
The two-dimensional melting transition of the hard sphere system has been 
an unsolved problem for many years\footnote{A 
review of two-dimensional melting is 
given in \cite{STRAND}.}. 
There are several theoretical approaches for the
description  \cite{KTHNY,CHUI}
and many numerical investigations have been done to analyze  
kind and order of this transition. The order-disorder transition 
was first seen in a computer study 
by  Alder and Wainwright \cite{ALDWAI}. 
They performed a numerical simulation
with the molecular dynamics algorithm
(constant $NVE$ simulations). 
Recent Monte Carlo investigations  were done in the $NVT$ 
\cite{ZOLCHE,WEMABI} and  $NpT$ ensemble \cite{LEESTR,FEALST}.
Unfortunately, these simulations gave non-unique results. 

One of the main problems of  Monte Carlo simulations 
is the autocorrelation in the sequence of generated 
configurations. This leads to a
drastical increase of computer time when the system 
is in  vicinity of the phase transition point. 
Roughly speaking, the autocorrelation time scales like
$\tau \sim \xi^z$, where $\xi$ is the correlation length
and $z$ denotes the dynamical  critical exponent.
This phenomenon  is called  critical
slowing down. In particular, conventional local algorithms
such as the Metropolis algorithm are 
affected by this problem ($z \approx 2$).
For some simple spin systems, a 
dramatic reduction of critical 
slowing down ($z \approx 0$)  can be achieved by the cluster
algorithm \cite{CLUSTER1,CLUSTER2}. 
An attempt to  implement such an algorithm
for hard core systems was proposed \cite{DREKRA}
\footnote{A lattice version of this algorithm is given in 
\cite{HERBLO}.}, but several 
complications arose near the critical point. Also, no
practical investigations of autocorrelation times were made.

This paper presents
an improved Metropolis algorithm, 
applicable to arbitrary hard core systems in any dimensions. For 
that purpose the usual local Metropolis 
algorithm \cite{METRO} is replaced by
a non-local {\it chain Metropolis algorithm}. In the special
case of hard disks in two dimensions we 
show that this leads to a  remarkable reduction of autocorrelation
times. 
 
\section{The algorithm}
\begin{figure}
\centerline{\epsfxsize=8.5cm
\epsfbox{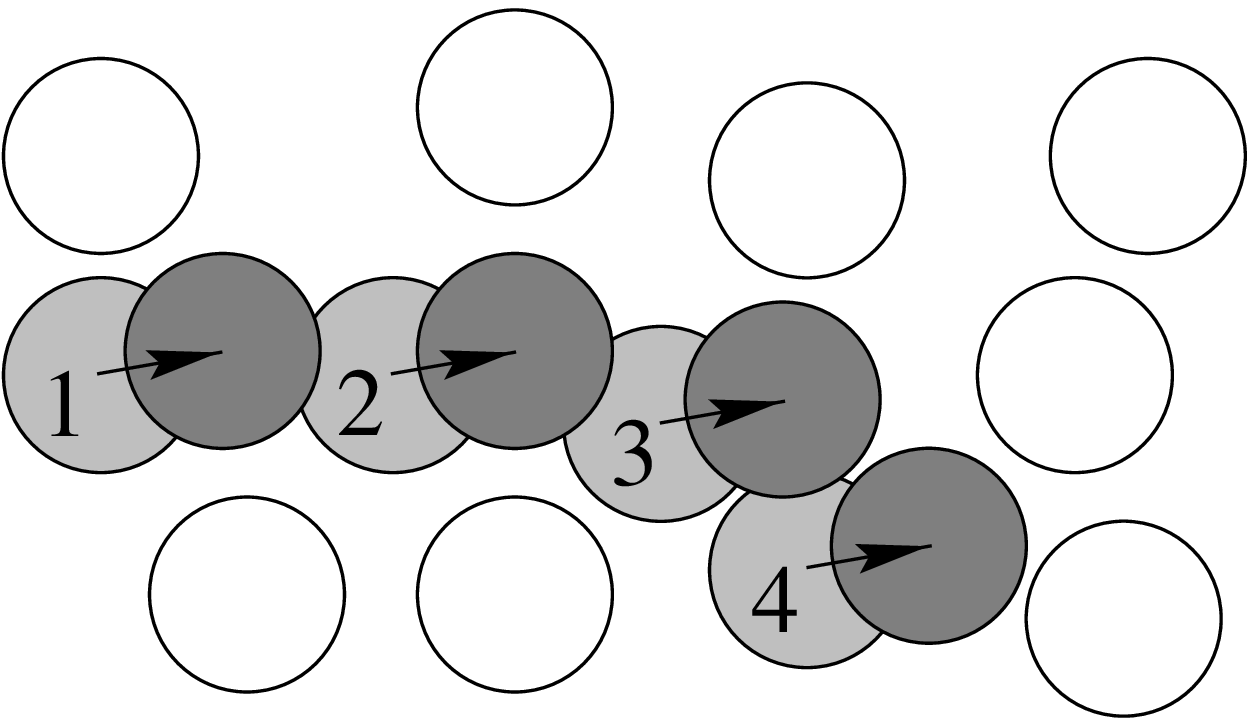}}
{\caption{\label{fig_schema}
Schematic illustration of an improved Metropolis step  for a hard 
disk system. Light grey disks indicate the initial positions,
while dark grey disks show the final ones. The process started
with disk `1' and ended at `4'.
}}
\end{figure}

Normally a Metropolis step for an 
$n$-dimensional hard core system
consists of the following operations:
\begin{enumerate}
\item
Select a particle $i$  from the system (this can be
done randomly with equal probability   or by 
sweeping over the whole system). 
\item
Displace
the particle from its position $\vec{r}_i$ with a uniform probability
(i.e.\ $\sim dV$)
to any point inside a cube\footnote{The region and
the distribution of
the new position could  be of other types.
For simplicity, we just consider  the case in the text.} 
of size $(2 \epsilon)^n$, centered at 
$\vec{r}_i$.
\item 
Accept the step, if the hard cores do not overlap.
\end{enumerate}
The idea of the improved Metropolis algorithm is to
displace chains of particles, thus enhancing the efficiency.
For simplicity and to decrease random number
generations, all particles of a chain
are moved in the same direction
with equal displacements.
Thus the improved (chain) Metropolis step  
is as follows:
\begin{enumerate}
\item
Select a particle $i$ randomly with equal probability   from the system.
\item
Select with  uniform probability a vector
$\delta \vec{r}$ of a cube of size $(2 \epsilon)^n$, centered at the origin.
\item
\label{OpDisplace}
Displace the particle from its position: 
$\vec{r}_i \rightarrow \vec{r}_i\,' = \vec{r}_i + \delta \vec{r}$.
\item
If the moved particle has an overlap with 
\begin{itemize}
\item[-]
no other particle, accept the step.
\item[-]
one particle, go back to operation \ref{OpDisplace} and displace the new
particle (i.e.\ the particle which has the overlap with the old one) 
by $\delta \vec{r}$.
\item[-]
two or more particles,  reject the step and
start from the beginning.
\end{itemize}
\end{enumerate}
An illustration of this process is given in figure \ref{fig_schema}.
Obviously, this updating process is ergodic and
satisfies the detailed balance condition. 
The algorithm is constructed in such a way that it produces
long chains in ordered areas (i.e.\ when the position correlation
length is large), while disordered regions lead
to short chains. Consequently the new algorithm spends more time
in the ordered domains. The stability of these 
ordered regions is responsible for the large autocorrelation
in the two-dimensional hard disk system,
so that the new non-local algorithm can lead to a decrease in the
autocorrelation times.

Moreover, the updating process
can be modified in such a way that the maximum number
of moved particles  is restricted
to $k_{\mathrm{max}}$, i.e.\ the step is rejected if the number of
particles in the chain exceeds $k_{\mathrm{max}}$.
The value of the step size $\epsilon$  for the improved
Metropolis algorithm
as well as for the original algorithm have to be tuned in such
a way that autocorrelation times are minimized.

\section{Autocorrelation times}
\begin{figure}
\begin{center}
\mbox{\epsfxsize=9.2cm
\epsfbox{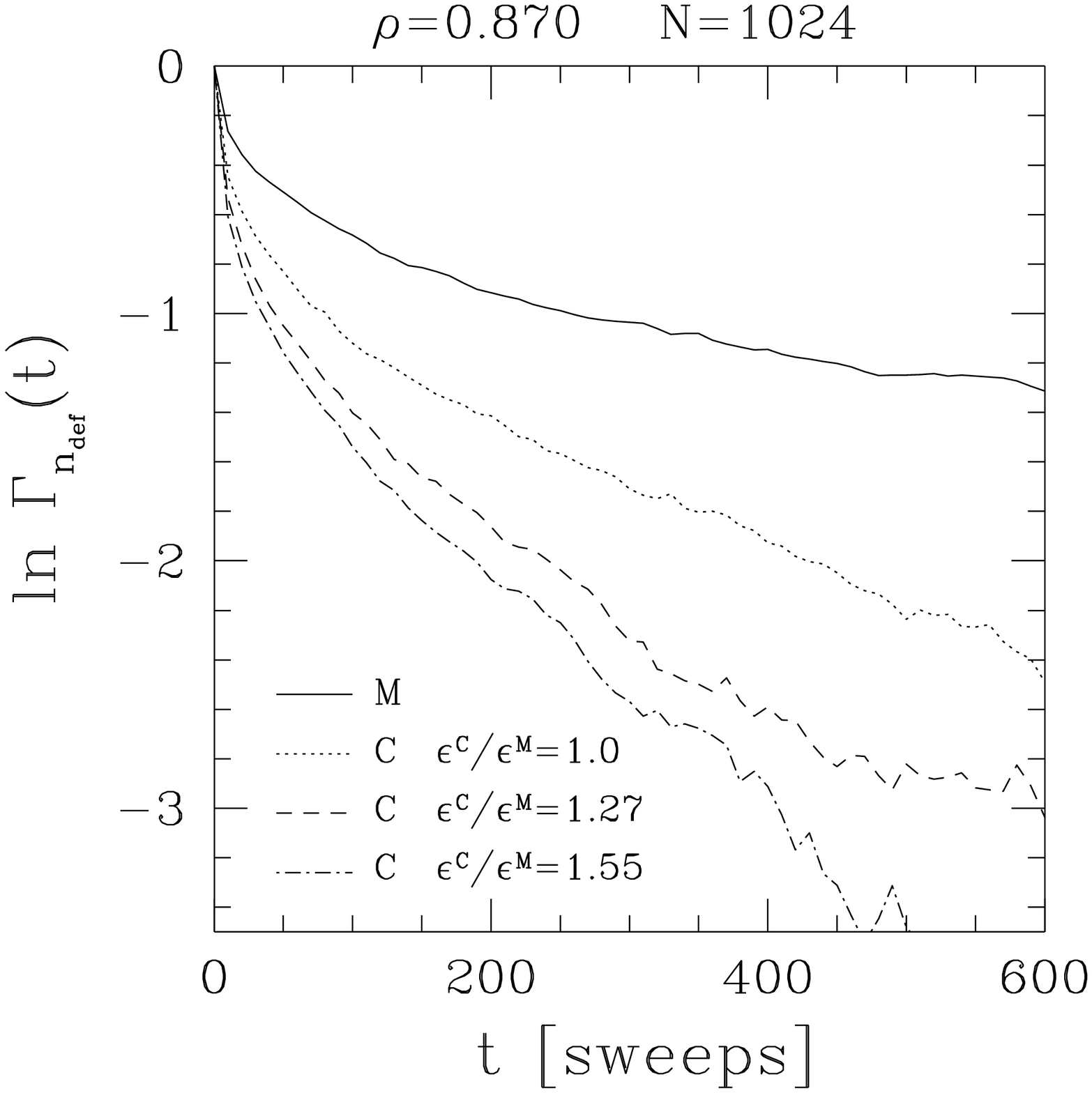}}
{\caption{\label{fig_autocor}
Autocorrelation functions for the fractional number of defects.
The hard disk system contained  $1024$ particles
at $\rho=0.870$. `M' and `C' denote the original and
the chain Metropolis updating scheme,  respectively. 
 The step sizes 
are given in units of the step size for the Metropolis algorithm 
$\epsilon^{\mathrm{M}}$. $\epsilon^{\mathrm{M}}$ 
was chosen in such a way that the acceptance 
rate was about $50\%$, which yields approximately 
the optimal value, i.e.\ minimal autocorrelation times 
(slight deviations in $\epsilon^{\mathrm{M}}$ 
from the optimal value lead only to non-essential changes).
}}
\end{center}
\end{figure}

In order to test the improvement of the chain Metropolis algorithm,
we studied the two-dimensional hard disk system in the $NVT$ ensemble
in the vicinity of the transition density 
$\rho_{\mathrm{c}}\approx0.898$ \cite{WEMABI},
where $\rho$ is given reduced units.
We used a square box with periodic boundary conditions and
measured the normalized autocorrelation function $\Gamma_{\mathrm{O}}(t)$
for an observable $O$ which is defined as
\begin{equation}
\Gamma_{\mathrm{O}}(t)=\frac{\langle O(0)\,O(t) \rangle - 
\langle O \rangle ^2}{\langle O^2 \rangle - \langle O \rangle ^2} \ ,
\end{equation}
where $t$ indicates the fictitious Monte Carlo time.
As an example,  figure \ref{fig_autocor} shows
the autocorrelation function for
the fractional number of defects
\begin{equation}
n_{\mathrm{def}}=\sum_{i\ne6}N_i/N \, , 
\end{equation}
where $N$ is the total number of 
particles and $N_i$ the number particles 
with $i$ neighbours determined by the Voronoi construction
\cite{VORONOI}.  The observable
$n_{\mathrm{def}}$ characterizes the defect structure of the two-dimensional
system. For the 
conventional Metropolis algorithm,  particles were selected
sequentially in order to cut down the amount of random number generation.
A sweep corresponds to one trial  of changing the 
coordinates for every particle.
For the chain Metropolis algorithm, particles 
have to be chosen randomly. 
In this case, we define a `sweep'  as $N$ different
trials to move chains of particles.
Obviously, for the same step size $\epsilon$
($\epsilon^{\mathrm{C}}/\epsilon^{\mathrm{M}}=1.0$),
the improved update scheme leads to a faster
decrease of the autocorrelation function. This can be
easily explained by the fact that every  displacement
of a particle $i$ 
accepted by the conventional Metropolis step
will also be  accepted by the improved step.
An increase of the
step size $\epsilon$ for the chain Metropolis algorithm 
leads to a faster decrease
of the autocorrelation function. However, for large 
values of $\epsilon$ a further increase results in
an increase of the autocorrelation function.

A realistic comparison of the new and the conventional 
Metropolis algorithm has to be done with autocorrelation times 
measured in computer time, since a step of a
chain of particles is more time-consuming. Therefore, 
we calculated the integrated autocorrelation time 
\begin{equation}
\tau_{\mathrm{int,O}}=\frac{1}{2} + \sum_{t=1}^\infty \Gamma_{\mathrm{O}}(t) 
\end{equation} 
in sweeps and in CPU time. Again, the values for the observable
$n_{\mathrm{def}}$  are collected
in table \ref{table1}. The  
new algorithm leads to a substantial decrease in the 
integrated autocorrelation time. The improvement is better for large
$k_{\mathrm{max}}$.
For the largest step size $\epsilon^{\mathrm{C}}$,
the autocorrelation time measured in  CPU time seems to increase.
It is not clear if this increase already
sets in or if it is just a statistical
effect. However, the exact  minimum of $\tau(\epsilon)$ 
is not very important,
since it is flat.
\begin{table}[t]
{\caption{ \label{table1} 
Integrated autocorrelation times of the improved Metropolis algorithm 
for the fractional number of defects  $n_{\mathrm{def}}$. 
The acceptance rate for the conventional Metropolis algorithm
was $53$--$55\%$, which corresponds approximately to the optimal
value of  $\epsilon^{\mathrm{M}}$ 
(which minimizes the integrated autocorrelation time in sweeps
as well as in CPU time).
In this case the particles were selected sequentially
to cut down the amount of random number generation.
}}
\begin{center}
\vspace*{4.0mm}
\begin{tabular}{lcr@{.}lcc*{2}{r@{.}l}}
\hline
\hline 
\vspace*{-5.0mm} \\
\multicolumn{1}{c}{$\rho$}  & $N$ & 
\multicolumn{2}{c}{$\epsilon^{\mathrm{C}}/\epsilon^{\mathrm{M}}$} &
$k_{\mathrm{max}}$  & $ \begin{array}{cc} 
\mbox{acceptance} \\ \mbox{rate} \end{array} $& 
\multicolumn{2}{l}{$\begin{array}{cc}
\tau^{\mathrm{C}}/\tau^{\mathrm{M}} \\ \mbox{\footnotesize [sweeps]}
\end{array}$}  &
\multicolumn{2}{l}{$\begin{array}{cc}
\tau^{\mathrm{C}}/\tau^{\mathrm{M}} \\ \mbox{\footnotesize [CPU time]}
\end{array}$} 
\vspace*{1.0mm} \\
\hline
0.870& 1024 & 1&00 & $N$ & 88\% & \phantom{00} 0&24 & \phantom{00} 0&34 \\  
0.870& 1024 & 1&27 & $N$ & 81\% & 0&13 & 0&20 \\  
0.870& 1024 & 1&27 &  6  & 80\% & 0&14 & 0&22 \\  
0.870& 1024 & 1&27 &  3  & 74\% & 0&23 & 0&32 \\  
0.870& 1024 & 1&55 & $N$ & 73\% & 0&11 & 0&18 \\  
0.870& 1024 & 1&91 & $N$ & 63\% & 0&069 &0&11 \\ 
0.870& 4096 & 1&91 & $N$ & 63\% & 0&14  &0&31 \\  
0.870& 1024 & 2&27 & $N$ & 53\% & 0&070 &0&12 \\ 
\hline
0.898& 1024 & 1&00 & $N$ & 86\% & 0&18 & 0&40 \\ 
0.898& 1024 & 1&27 & $N$ & 79\% & 0&12 & 0&29 \\  
0.898& 1024 & 1&27 &  6  & 78\% & 0&14 & 0&34 \\  
0.898& 1024 & 1&27 &  3  & 72\% & 0&24 & 0&53 \\  
0.898& 1024 & 1&55 & $N$ & 70\% & 0&11 & 0&27 \\  
0.898& 4096 & 1&55 & $N$ & 70\% & 0&14 & 0&32 \\ 
0.898& 1024 & 1&91 & $N$ & 59\% & 0&12 & 0&29 \\  
\hline
\hline
\end{tabular} 
\end{center}
\end{table}

In addition to the fractional number of defects, we studied the 
second moment of the bond orientational order parameter 
\begin{equation}
{\psi_6}^2=\, \left < \left | 
\frac{1}{N} \sum_j \frac{1}{N_i} \sum_{k=1}^{N_i} 
\exp(6{\mathrm{i}}\,\theta_{kj}) \right | ^2 \right > \ ,
\end{equation} 
where the first sum is over all
particles, the sum on $k$ is over the $N_i$ neighbours of the particle
$j$, and $\theta_{kj}$ is the angle  between
the particles $k$ and $j$ and an arbitrary but fixed  reference axis. 
The bond orientational order parameter is, in contrast to the defect density,
a global quantity and describes the  orientational order. The values
of ${\psi_6}^2$ lie between $0$ and $1$, where the latter 
corresponds to  a perfect crystalline structure.

Integrated  autocorrelation times for the
second moment of the bond orientational order parameter
are larger. 
For example, for the $4096$ particle system at $\rho=0.870$,  
$\tau^{\mathrm{C}}_{{\psi_6}^2}/\tau^{\mathrm{C}}_{n_{\mathrm{def}}}\approx 3.5$.
Indeed, the stability of large  crystalline regions
prevents large changes  of  the global quantity ${\psi_6}^2$.
Therefore,  the determination is more time-consuming, and more
attention  was given to the defect density. Furthermore, the investigations
of ${\psi_6}^2$ point to similar conclusions as those of $n_{\mathrm{def}}$. 
For instance
at $\rho=0.870$ and $\epsilon^{\mathrm{C}}/\epsilon^{\mathrm{M}}=1.55$,
$\tau^{\mathrm{C}}/\tau^{\mathrm{M}}=0.15$ (in computer time)
for  $1024$ particles,
and $\tau^{\mathrm{C}}/\tau^{\mathrm{M}}=0.32$ for $4096$ particles
were obtained for ${\psi_6}^2$.
Some runs at $\rho=0.9$ and 
$\epsilon^{\mathrm{C}}/\epsilon^{\mathrm{M}}=2.0$ 
with $4096$ and $16384$ hard disks were also performed, yielding
an estimate of $\tau^{\mathrm{C}}/\tau^{\mathrm{M}} \approx 0.2$.

Investigations of 
the computer time used  showed that  it
approximately scales  with the number of
particles (for systems
with more than 1024 particles)
for both densities. This is the expected behaviour for the local
Metropolis algorithm.
In the case of the improved algorithm it indicates
that large chains  play no significant role. 
We checked this by studying the chain length in systems with
$4096$ particles at both densities. Indeed, chains with
more than $16$ particles are very rare.
Therefore, one
might expect that the chain Metropolis algorithm
does not lead to an improvement of the dynamical critical exponent.
Integrated autocorrelation times for the $4096$ particle system at 
the critical density
seem to confirm this assumption, but are affected by large statistical errors.
Hence it appears that critical slowing down is not reduced.

\section{Modification}
\begin{figure}
\begin{center}
\mbox{\epsfxsize=7.6cm
\epsfbox{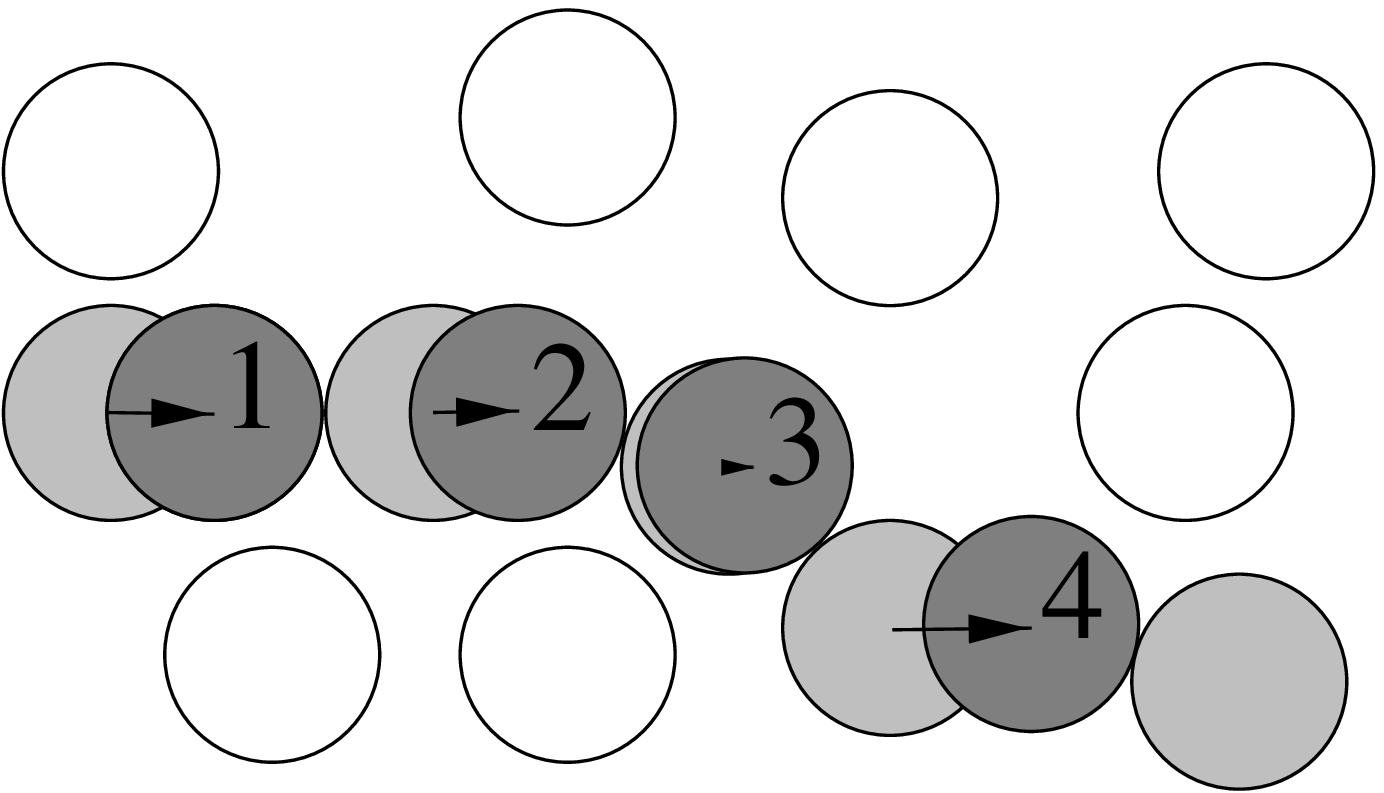}}
\hspace{2.0eM}
\mbox{\epsfxsize=5.1cm
\epsfbox{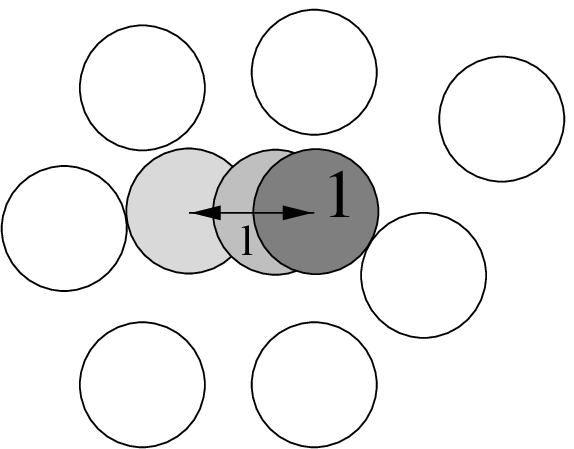}}
{\caption{\label{fig_schema1}
The left figure illustrates the second updating scheme. 
A particle (`1') is moved in a randomly chosen direction until 
it collides with another one (`2'). The second particle 
is moved into the same direction up to the next collision.
This process is stopped after $k$  particles.
The last particle is randomly placed between its initial position
and the collision point in $\vec{v}$ direction.
The right figure illustrates the length $l_{\mathrm{start}}$, which is
marked by an arrow. The disk in the middle is the 
initial position of particle `1', the right disk shows the final
position, while the left one corresponds to the collision point
in $-\vec{v}$ direction.
The length $l_{\mathrm{start}}$ is now the distance between the final
position and the collision
point in the negative $\vec{v}$ direction.
}}
\end{center}
\end{figure}

A further improvement can  perhaps be  achieved by an
updating scheme which enables the interaction of
particles over large distances. 
Since the algorithm described above stops if more 
than two particles overlap, this situation
must be avoided. Therefore, each particle
is moved in a chosen direction just to the point
where it collides with another particle. For simplicity,
we consider the case where all particles are 
moved in the same direction $\vec{v}$.
With the number of particles $k$ in the chain 
chosen arbitrarily, this translates into:
\begin{enumerate}
\item
Select randomly  a
number $k$ (the number of particles to move) 
between 1 and $k_{\mathrm{max}}$. Select 
randomly with equal probability
a particle $i$ (the first particle in the chain)
from the whole system.

\item
Choose a vector $\vec{v}$ of the surface of a
unit sphere, i.e.\ with a probability  $\sim  d\Omega$.

\item
\label{OpMove}
Move the particle into the direction of $\vec{v}$ until a collision
with another particle occurs. Add this to the chain.

\item
If the number of particles in the chain (including the new one) is
\begin{itemize}
\item[-]
lower than $k$, go back to operation \ref{OpMove} and move the
new particle in the same direction.

\item[-]
equal to $k$, place the particle with uniform
probability between 
its initial position and the 
collision point   (in $\vec{v}$ direction), i.e.\
$\vec{r}_k \rightarrow \vec{r}_k\,'=\vec{r}_k+ 
r(\vec{r}_k^{\mathrm{\ coll}} - \vec{r}_k)$,
where $r$ denotes a uniform distributed random number between 0 and 1.

\end{itemize}
\item
\label{OpAcceptReject}
Accept the new configuration (the displacement of the whole chain)
with probability
$\min(1,l_{\mathrm{end}}/l_{\mathrm{start}})$. 
$l_{\mathrm{end}}=| \vec{r}_k^{\mathrm{ \ coll}} - \vec{r}_k |$ is the
length between the initial position and the collision 
point of the last particle;  
$l_{\mathrm{start}}$ is the 
length from the final position of the first particle 
to the collision point in $-\vec{v}$ direction.
\end{enumerate}
This updating step and the meaning of $l_{\mathrm{start}}$
is explained schematically in figure~\ref{fig_schema1}.
The new accept/reject step (operation \ref{OpAcceptReject}) is necessary to
fulfill the detailed balance condition.
The case  $k=1$ is realized by placing 
the particle  randomly between the
collision points in positive  and   negative $\vec{v}$ direction. 
This step is always accepted. The value of $k_{\mathrm{max}}$ as well as the
distribution of $k$ between $1$ and $k_{\mathrm{max}}$
(which is not necessarily 
chosen  with equal probability)
have to be tuned in such a way that autocorrelation times are
minimal.

The advantage of this updating scheme is a  high and 
length-independent probability to accept 
a step, which perhaps  can counter critical slowing down.
On the other hand,  a displacement of a chain with $k$ particles
 is more time-consuming  due to the additional 
calculations of the collision points
than a comparable step in the first (improved) updating  scheme. Therefore, 
a detailed investigation of this algorithm, in particular
for the dynamical critical exponent,
is necessary in order to test the performance. So far 
only a few runs have been performed to 
ensure that it yields the correct expectation values
and to get a rough estimate of the integrated autocorrelation time
for ${\psi_6}^2$.
For $16384$ hard disks at $\rho=0.9$,  
the ratio of the integrated autocorrelation time (in sweeps) 
of this updating scheme to those of the first scheme, 
$\tau^{\mathrm{C'}}/\tau^{\rm{C}}$,
was about $0.22$, while the value measured in CPU time strongly
depends on the implementation of the algorithms.

\section{Conclusions}
In summary, we presented a non-local Metropolis algorithm for 
arbitrary hard core systems in any dimension.
We investigated the two-dimensional hard disk system for a detailed
comparison between the conventional local and improved non-local
algorithm. The Monte Carlo simulations showed that the new 
algorithm leads to an essential 
reduction of integrated autocorrelation times for the fractional
number of defects as well as the global bond orientational parameter. 
An additional modification which perhaps leads to a further 
improvement was presented, but no investigations were made.

It is hoped that this algorithm  will help to resolve the type and order 
of the phase transition in the two-dimensional 
hard disk system.   Work along this line is in progress.

\begin{ack}
Critical comments by Joachim Hein, Karl Jansen
and Jacques Mainville are
gratefully acknowledged. We thank Harro Hahn for useful
discussions.
\end{ack}

\end{document}